\documentclass[aps,prd,amsmath,floats,floatfix, twocolumn,
superscriptaddress,nofootinbib,showpacs]{revtex4-1}

\usepackage[colorlinks, pdfborder={0 0 0}, plainpages=false]{hyperref}
\usepackage{graphicx}
\usepackage{xspace}
\usepackage[usenames,dvipsnames]{color}
\usepackage{amssymb}

\newcommand{\roughly}{\mathchar"5218\relax} 

\newcommand{\run}[1]{\ensuremath{S^{++}_{#1}}}

\newcommand{\Caltech}{\affiliation{Theoretical Astrophysics 350-17,
    California Institute of Technology, Pasadena, CA 91125, USA}}
\newcommand{\Cornell}{\affiliation{Center for Radiophysics and Space
    Research, Cornell University, Ithaca, New York 14853, USA}}
\newcommand{\CITA}{\affiliation{Canadian Institute for Theoretical
    Astrophysics, 60 St.~George Street, University of Toronto,
    Toronto, ON M5S 3H8, Canada}} %
\newcommand{\CIFAR}{\affiliation{Canadian Institute for Advanced Research, 180 Dundas St.~West, Toronto, ON M5G 1Z8, Canada}} %
\newcommand{\GWPAC}{\affiliation{Gravitational Wave Physics and
    Astronomy Center, California State University Fullerton,
    Fullerton, California 92834, USA}} %
\newcommand{\Oberlin}{\affiliation{Department of Physics and Astronomy,
    Oberlin College, Oberlin, Ohio 44074, USA}} %
\newcommand{\Princeton}{\affiliation{Department of Physics,
    Princeton University, Jadwin Hall, Princeton, NJ 08544, USA}} %

\begin{document}

\title{
Nearly extremal apparent horizons in simulations of merging black holes 
}

\author{Geoffrey~Lovelace}\GWPAC\Caltech
\author{Mark~A.~Scheel}\Caltech
\author{Robert~Owen}\Oberlin
\author{Matthew~Giesler}\Caltech\GWPAC
\author{Reza~Katebi}\GWPAC
\author{B\'{e}la Szil\'{a}gyi}\Caltech
\author{Tony~Chu}\Princeton\CITA
\author{Nicholas~Demos}\GWPAC
\author{Daniel~A.~Hemberger}\Caltech
\author{Lawrence~E.~Kidder}\Cornell
\author{Harald~P.~Pfeiffer}\CITA\CIFAR
\author{Nousha~Afshari}\GWPAC

\date{\today}

\begin{abstract}
The spin angular momentum $S$ of an isolated Kerr black hole is bounded by
the surface area $A$ of its apparent horizon: $8\pi S \le A$, with 
equality for extremal black holes.
In this paper, we explore the extremality of individual and common 
apparent horizons for merging, rapidly spinning binary black holes. 
We consider simulations of merging black holes with equal masses $M$ and 
initial spin angular momenta aligned with the orbital angular momentum, 
including new simulations with spin magnitudes up to $S/M^2 = 0.994$. 
We measure the area and (using approximate 
Killing vectors) the spin on the individual and common apparent horizons, 
finding that the inequality $8\pi S < A$ is satisfied in all cases 
but is very close to equality on the common apparent horizon at the 
instant it first appears. We also evaluate the Booth-Fairhurst 
extremality, whose value for a given apparent horizon 
depends on the scaling of the horizon's null normal vectors. 
In particular, 
we introduce a gauge-invariant lower bound on the extremality by computing 
the smallest value that Booth and Fairhurst's extremality parameter can take 
for any scaling. Using this lower bound, we conclude that the common horizons 
are at least moderately close to extremal just after they appear.  
Finally, following 
Lovelace et al.~(2008),
we construct quasiequilibrium 
binary-black-hole initial data 
with ``overspun'' marginally trapped surfaces with $8\pi S > A$.
We show 
that the overspun surfaces are indeed superextremal: 
our lower bound on their Booth-Fairhurst 
extremality exceeds unity.
However, we confirm that these superextremal surfaces are
always surrounded by marginally outer trapped surfaces
(i.e., by apparent horizons) with $8\pi S<A$.  The 
extremality lower bound on the enclosing apparent horizon is always 
less than unity but can exceed the value for an
extremal Kerr black hole.
\end{abstract}


\maketitle

\section{Introduction}

\subsection{Motivation and background}
In the decade following 
Pretorious's breakthrough simulation~\cite{Pretorius2005a}, 
and the development of the moving puncture 
technique~\cite{Campanelli2006a, Baker2006a},
several research groups have made 
great strides toward simulating merging binary black holes (BBHs)
with a variety of mass ratios and spins (for recent reviews, see, 
e.g., ~\cite{Centrella:2010,Pfeiffer:2012pc,Hannam:2013pra,Tiec:2014lba}).
These BBH simulations reveal 
not only the emitted gravitational waves but also 
the behavior of the strongly warped, highly dynamical spacetime near the holes' 
horizons. For example, a number of research groups have 
explored the properties (mass, spin, and recoil 
velocity)
of the final, remnant hole in BBH mergers~\cite{Campanelli2007a,
Gonzalez2007b,Rezzolla:2007xa,Tichy2008,Barausse2009,Reisswig:2009vc,
Lousto:2011kp,Barausse:2012qz,Hemberger:2013hsa,London:2014cma,
Lousto:2013wta,Healy:2014yta}.
These studies typically follow a 
``scattering matrix'' approach, understanding the nonlinear dynamics by 
exploring how different initial BBH configurations produce different 
remnant Kerr black holes. 
Some studies have explored the highly nonlinear dynamics of the strongly 
curved spacetime during the merger itself; these include 
recent work using tendex and vortex lines 
(analogous to electric and magnetic field lines) 
to visualize the curvature of 
simulated spacetimes~\cite{OwenEtAl:2011} and recent work exploring 
how the remnant properties are imprinted on the emitted gravitational 
waves~\cite{Healy:2014eua}.

In this paper, we consider the \textit{extremality} of dynamical black 
holes in numerical spacetimes. 
A single Kerr black hole with spin magnitude $S$, horizon area $A$, and 
mass $M$ obeys 
the inequalities
\begin{eqnarray}
8\pi S & \le & A,\label{ineq:spinArea}\\
S & \le & M^2\label{ineq:spinMass},
\end{eqnarray}
or
\begin{eqnarray}
\zeta \equiv \frac{8\pi S}{A} & \le & 1,\label{ineq:zeta}\\
\chi \equiv \frac{S}{M^2} & \le & 1.\label{ineq:chi}
\end{eqnarray} 
For a Kerr black hole, both $\chi$ and $\zeta$ can be interpreted as 
measures of the hole's extremality; 
a Kerr hole is nearly extremal if $\chi \approx 1$ and $\zeta \approx 1$. 

The actual spins and extremalities of astrophysical black holes are 
uncertain, but 
there is observational evidence that nearly extremal black holes could
exist. For instance, recent measurements using both 
continuum fitting and x-ray reflection fitting suggest that 
Cygnus X-1 (the first black hole discovered) is nearly 
extremal~\cite{Gou:2011nq,Fabian:2012kv,Gou:2014una}; 
there are also measurements of nearly extremal spin in other 
stellar-mass black holes (such as in x-ray binaries  
GRS 1915+105~\cite{McClintockEtAl:2006} and GX 339-4~\cite{Miller:2009cw}) 
and in supermassive black holes (e.g.~Swift J0501.9-3239~\cite{Walton:2012aw}).
(For reviews of black-hole spin measurements, see, e.g., 
Refs.~\cite{McClintock:2013vwa,Reynolds:2013qqa}.)
Therefore, BBHs targeted by gravitational-wave detectors could contain
nearly extremal black holes; 
this has 
motivated previous and ongoing 
efforts to simulate BBH mergers with nearly extremal 
spins~\cite{Rezzolla:2007xa,Lovelace2008,HannamEtAl:2010,
MarronettiEtal:2008,DainEtAl:2008,Lovelace:2010ne,Lovelace:2011nu,
Mroue:2013PRL,Ruchlin:2014zva}. 

Can such numerical
simulations of merging, nearly extremal black holes contain
superextremal horizons? Answering this question requires a
generalization of extremality from
the Kerr solution, 
preferably one that can be measured quasilocally (e.g.~on apparent horizons). 
For exact axisymmetry, an unambiguous spin $S$ (the 
Komar spin angular momentum~\cite{Komar:1958wp}) can 
be defined on the apparent horizon; for this spin
measure, inequality~(\ref{ineq:spinArea}) 
has been proven to hold for spacetimes 
satisfying the Einstein equations with non-negative cosmological constant and 
with non-electromagnetic matter fields obeying the dominant energy 
condition~\cite{Jaramillo:2011pg}. (See Ref.~\cite{Dain:2011mv}
and the references therein for a review of geometric inequalities that 
have been proven for axisymmetric black holes.) 

BBHs, in contrast, do not
have exact symmetries to facilitate definitions of conserved mass and
angular momentum. In Ref.~\cite{BoothFairhurst:2008}, Booth and Fairhurst 
argue that extremality bounds such as inequalities~(\ref{ineq:spinArea}) 
or~(\ref{ineq:spinMass}) are not well posed, since there are no symmetries to 
define mass and angular momentum. 
Booth and Fairhurst propose an alternative 
extremality parameter $e$ that does not require approximate axial 
symmetry. In practice, numerical relativists typically measure 
black-hole spin angular momentum $\chi$ (and thus the extremality $\zeta$) 
in terms of approximate 
symmetries characterized by approximate Killing vectors.

\subsection{Overview of results}

In this paper, we explore two alternative measures of extremality in
numerical simulations of merging black holes. First, we explore
whether inequality~(\ref{ineq:spinArea}) is violated on apparent
horizons of area $A$, evaluating the spin $S$ of the apparent horizon
using approximate Killing vectors, which identify the best approximate
symmetries available.  We confine our attention to a set of BBH 
simulations performed using the Spectral Einstein Code 
(SpEC)~\cite{SpECwebsite}. Each simulation has
equal masses and equal, aligned spins. 
Even with spins as high as
$S/M^2 = 0.994$, the highest yet simulated, we observe no 
violation\footnote{Note, however, that Bode, Laguna,
and Matzner have found that inequality~(\ref{ineq:spinArea}) can be
violated by accreting matter with negative energy density (in the
form of constraint violation) onto a black
hole~\cite{Bode:2011xz}.}
of
inequality~(\ref{ineq:spinArea}). When a common horizon first forms, 
however, $A$ and $8\pi S$ can be very close to each other: for a BBH 
with an initial spin of $\chi=0.994$, we find $\zeta$ as high as   
$0.997$.

Second, we revisit the Booth-Fairhurst extremality $e$, whose value
depends on a choice of scaling for the ingoing and outgoing null
normals to the horizon. We define a gauge-invariant lower bound of the
extremality, $e_0$, as the smallest value $e$ can take for any scaling
of the null vectors. Whatever gauge-fixing prescription might be used,
the extremality $e$ will always be larger than or equal to $e_0$. 
The lower bound $e_0$ can be moderately large: for
  merging holes with initial spins $\chi=0.994$, $e_0$ of the common
  apparent horizon, just after it appears, is comparable to $e_0$ of a
  Kerr black hole with a spin of $\chi \approx 0.97$. This suggests
  that these newly formed common apparent horizons must be at least
  moderately close to extremality.

Third, we construct binary-black-hole initial data with
superextremal marginally outer trapped surfaces
that violate Eq.~(\ref{ineq:spinArea}).
As in Ref.~\cite{Lovelace2008}, a larger marginally outer trapped
surface (the apparent horizon) which satisfies (\ref{ineq:spinArea})
always encloses these overspun surfaces. Our lower bound of extremality $e_0$ 
exceeds unity 
on the inner, overspun surfaces, verifying that they are in fact 
superextremal. 
On the apparent horizons, we find that $e_0$ is less than unity 
but can exceed the value of $e_0$ for an extremal 
Kerr black hole.

The rest of this paper is organized as follows. In Sec.~\ref{sec:techniques}, 
we summarize our numerical methods and our methods for 
measuring extremality, including our lower bound for 
the extremality of an apparent horizon. In Sec.~\ref{sec:results}, we 
present results for the extremalities of apparent horizons in BBH 
simulations with merging, rapidly spinning black holes. We
conclude in Sec.~\ref{sec:conclusions}.

\section{Techniques}\label{sec:techniques}
In this section, we describe our numerical techniques. 
First, 
we discuss our method, based on approximate Killing vectors 
(Sec.~\ref{sec:AKV}), for measuring spins on apparent horizons. 
Then, we describe other methods for directly measuring 
extremality, introducing a new lower bound of the extremality of an apparent 
horizon (Sec.~\ref{sec:extremality}). 
We conclude this section by 
summarizing the methods we use to 
simulate the merging black 
holes that we will examine (Sec.~\ref{sec:simulationMethods}). 

\subsection{Defining spin by approximate Killing vectors}\label{sec:AKV}

The standard method for computing spin in numerical relativity is the
following integral, carried out on the apparent-horizon
2-surface~\cite{BrownYork1993, Ashtekar2001, Ashtekar2003}:
\begin{equation}
S = \frac{1}{8 \pi} \oint \omega_B \phi^B dA.
\end{equation}
Here capital latin indices index the tangent bundle to the 2-surface within 
spacetime. The one-form $\omega_B$ physically represents a surface angular momentum density and mathematically represents a connection on the normal 
bundle, defined as
\begin{equation}
\omega_A = e^\mu_A n_\rho \nabla_\mu \ell^\rho,
\end{equation}
where $\vec n$ and $\vec \ell$ are (respectively) the ingoing and outgoing 
null normals to the 2-surface in spacetime, arrows denote spacetime 4-vectors, 
$\vec \nabla$ is the spacetime 
covariant derivative, and $e^\mu_A$ is the spacetime representation of a basis 
for the tangent bundle (a projector to the 2-surface). In practice, the 
form $\omega_A$ is usually computed from the extrinsic curvature of the 
spatial slice (e.g. Eq.~(A1) of Ref.~\cite{Lovelace2008}).

The vector field $\vec \phi$ on the 2-surface is a generator of
rotations, and this vector field encodes the directional nature of the
angular momentum. Here, we choose this vector field as in previous
papers, using methods of {\it approximate Killing
  vectors}~\cite{Dreyer2003, Cook2007, OwenThesis, Lovelace2008}. An
arbitrary 2-surface will generally not have symmetries, but this
method finds the vector field that comes {\it closest} to a symmetry
on an arbitrary 2-surface, in the sense of minimizing a residual of
Killing's equation with respect to variations in the space of smooth
vector fields. 

First, because Killing's equation implies that $\vec
\phi$ is divergence-free, we start with the condition
\begin{equation}
\phi^A = \epsilon^{AB} D_B z,
\end{equation}
for some smooth function $z$ on the 2-surface, where $D$ 
and $\epsilon$ are, respectively,
the covariant derivative and Levi-Civita tensor
intrinsic to the 2-surface.
This condition renders the computed 
spin invariant under the 
boost-gauge ambiguity, which is an ambiguity of the scaling
of the null normals. The null normals 
$\vec \ell$ and $\vec n$ are normalized only relative to one another, through 
the standard condition
\begin{equation}
\vec \ell \cdot \vec n = -1.
\end{equation}
This condition is preserved by the rescaling 
\begin{align}
\begin{split}
\vec \ell &\mapsto \exp(a) \vec \ell,\\
\vec n &\mapsto \exp(-a) \vec n,
\end{split}
\label{eq:boostgauge}
\end{align}
where $a$ is an arbitrary function on the 2-surface, which can be interpreted 
as a rapidity.
The surface angular momentum density transforms as
\begin{equation}
\omega_A \mapsto \omega_A + D_A a.
\label{eq:OmegaBoostTransform}
\end{equation}
Thus, under a rescaling of the null normals, the quasilocal angular momentum 
transforms as:
\begin{equation}
S \mapsto S + \oint \left(D_B a\right) \left(\epsilon^{BC} D_C z\right) \hspace{1mm} dA.
\end{equation}
An integration by parts and the condition of zero torsion then show 
that $S$ is unchanged by the boost transformation~(\ref{eq:boostgauge}).

A minimization problem for the integral of the square of the 
shear of $\vec \phi$ (the remainder of Killing's equation) then implies that 
$z$ must satisfy a generalized eigenproblem
\begin{equation}
D^4 z + \vec D \cdot \left( R \vec D z \right) = \lambda D^2 z,\label{e:eigenproblem}
\end{equation}
where $R$ is the intrinsic scalar curvature on the horizon and the eigenvalue 
$\lambda$ is related to the overall shear of the vector field $\vec \phi$. 
We solve a spectral representation of this generalized eigenproblem, find the eigenfunction 
$z$ with minimum eigenvalue\footnote{On highly deformed surfaces, the 
eigenfunction with smallest $\lambda$ is {\it not} the one associated with 
the black-hole spin. For example, on a ``peanut-shaped'' horizon immediately 
after binary black hole merger, the best approximate symmetry is usually a 
rotation about the axis connecting the centers of the progenitor holes.
In practice, we find 
that even on highly deformed and dynamical horizons there is one eigenfunction 
$z$ that gives a large value of the spin angular momentum, and a visual 
inspection shows that the rotation vector associated with this function points 
in the direction that one would intuitively associate with the rotation of 
the black hole.}, normalize it according to a prescription described in 
Ref.~\cite{Lovelace2008}, and compute the quasilocal spin angular 
momentum as
\begin{equation}
S = \frac{1}{8 \pi} \oint \omega_B \epsilon^{BC} D_C z \hspace{1mm} dA.
\end{equation}
Through an integration by parts, 
this equation can equivalently be written as
\begin{equation}\label{eq:spinDef}
S = \frac{1}{8 \pi} \oint z \Omega \hspace{1mm} dA,
\end{equation}
where $\Omega \equiv \epsilon^{AB} D_A \omega_B$ is a scalar curvature of the 
normal bundle of the 2-surface in spacetime. This quantity $\Omega$
is the same (up to 
a constant factor) as the imaginary part of the {\it complex curvature} defined 
in Ref.~\cite{Penrose1992}. Also, apart from correction terms that happen to 
vanish on isolated horizons, it is also equal to the imaginary part of the 
Weyl scalar $\Psi_2$ in a tetrad adapted to the 2-surface, or equivalently, 
to the normal-normal component of the magnetic component of the Weyl tensor, 
referred to as the {\it horizon vorticity} in Ref.~\cite{OwenEtAl:2011}.

This approach to defining black-hole spin springs from 
an assumption that the horizon is nearly axisymmetric, a condition that is 
blatantly violated in some of the cases we consider here. Thus, it is 
surprising and rather mysterious that we get any useful physics at all---such 
as the apparent satisfaction of the extremality bound $8 \pi S \le A$ 
(Sec.~\ref{sec:results}). 
Moreover, 
while the above construction provides an apparently reasonable implementation 
of the intuitive idea of ``approximate symmetry,'' it can be generalized 
rather significantly, providing alternative measures of spin that agree with 
this measure on axisymmetric surfaces, but which differ from it when the 
axisymmetry is broken. One such modified form is that in Ref.~\cite{Cook2007}, 
and a broader family is presented in Ref.~\cite{Owen2014}.

To make a clearer case 
for extremality bounds in highly-deformed horizons, we would like a measure 
of black-hole extremality that is independent of any assumptions of symmetry 
and any particular definition of black-hole spin.
We address this in the next subsection.

\subsection{Direct measures of horizon extremality}\label{sec:extremality}

On highly deformed horizons, standard measures of black-hole spin,
such as the one given in Eq.~(\ref{eq:spinDef}) and the surrounding 
discussion,
have a questionable
physical motivation. The root of the difficulty is the
need to define the vector field $\vec \phi$ on the surface. In the
absence of even approximate axisymmetry, there is no obviously
preferred vector field to choose. 
We continue to define spin using approximate Killing vectors, even though 
their use is no longer geometrically well-motivated.

In Ref.~\cite{BoothFairhurst:2008}, Booth and Fairhurst introduced a measure 
that avoids this issue. Their {\it extremality parameter} is 
\begin{equation}
e = \frac{1}{4 \pi} \oint \omega_B \omega^B \hspace{1mm} dA.
\end{equation}
Intuitively, this might be understood as a squared ``quasilocal spin
magnitude,'' though as was noted in Ref.~\cite{BoothFairhurst:2008},
because we are integrating over the continuous horizon 2-surface, it
might better be understood as including information from all of the
current multipole moments. We will specify this relationship more
precisely below.

The practical benefit of working with $e$ rather than $S$ is clear: it
removes the need for solving the eigenproblem given in 
Eq.~(\ref{e:eigenproblem}), 
and it
removes questions of interpreting the results when axisymmetry is
strongly broken. However, $e$ is not simply an ad-hoc quantity chosen
to avoid these practical issues. Reference~\cite{BoothFairhurst:2008}
shows that the value of $e$ is closely related to the question of
whether fully-trapped surfaces exist within the marginally-trapped
surface on which it might be computed, and that subextremal isolated
horizons (subextremal in the sense of having positive surface gravity) 
satisfy $e < 1$. 

In spite of these appealing features, there is one major practical drawback to
working with the Booth-Fairhurst extremality, related to the
issue of boost gauge described in Sec.~\ref{sec:AKV}. Under the
rescaling of $\vec \ell$ and $\vec n$ given in 
Eq.~(\ref{eq:boostgauge}),
$e$ transforms as 
\begin{equation}
e \mapsto e + 2 \oint \omega^B D_B a \hspace{1mm} dA + \oint \lvert\vec D a \rvert^2 dA.
\end{equation}
To talk about the extremality of the horizon in terms of the 
quantity $e$, one must fix a boost gauge.

The suggestion made in Ref.~\cite{BoothFairhurst:2008} for fixing the
boost gauge is a standard one in the dynamical horizon
literature. Because the dynamical horizon is a spacelike object, it is
natural to scale $\vec \ell$ and $\vec n$ such that the unit
spacelike normal to the 2-surface, $\hat s = \left(\vec \ell - \vec
n\right)/\sqrt{2}$, is tangent to the dynamical horizon. This
condition has many attractive mathematical
features~\cite{Ashtekar2003}, but it has one particularly worrisome
drawback in the numerical context: as a horizon settles down to that
of a Kerr black hole, it approaches a {\it null} surface in
spacetime. Boosting the spatial vector $\hat s$ to this asymptotically
null surface requires an arbitrarily large boost rapidity $a$. 
Reference~\cite{Schnetter2006} 
notes 
that the process of
calculating certain quantities in the boost gauge adapted to the dynamical
horizon becomes numerically ill-behaved as the horizon settles
down to Kerr.
Even more troubling, Ref.~\cite{Schnetter2006} finds 
inner horizons that are
spacelike in some regions and timelike in others. 

For these reasons, 
we have opted to take a different approach to fixing the boost gauge.
Our approach begins with a decomposition of the angular momentum surface 
density into two scalar potentials $\varpi$ and $\pi$,
\begin{equation}
\omega_A = \epsilon_A{}^B D_B \varpi + D_A \pi.
\end{equation}
Given any $\omega_A$, the two potentials can be computed (up to an irrelevant 
constant) by solving Poisson equations on the horizon 2-surface: 
\begin{eqnarray}
D^2 \pi &=& D^A \omega_A\\
D^2 \varpi &=& \epsilon^{AB} D_A \omega_B = \Omega.
\end{eqnarray}
Note that, by construction, the right-hand side of both of these equations 
averages to zero over the 2-surface, as is required for a solution to exist.

The two potentials $\varpi$ and $\pi$ are distinguished not only by parity 
considerations, but more importantly by how they behave under changes in 
boost gauge. 
Equation~(\ref{eq:boostgauge}) implies that the sources of the above 
Poisson equations transform as 
\begin{eqnarray}
D^A \omega_A &\mapsto& D^A \omega_A + D^2 a\\
\Omega &\mapsto& \Omega,
\end{eqnarray}
so that the potentials transform as 
\begin{eqnarray}
\pi &\mapsto& \pi - a\\
\varpi &\mapsto& \varpi.
\end{eqnarray}
In other words, given any $\omega_A$, one can always find the transformation 
into a preferred family of boost gauges in which
$\pi$ is constant,
and therefore 
\begin{equation}\label{eq:prefGauge}
\omega_A = \epsilon_A{}^B D_B \varpi, 
\end{equation}
where $\varpi = D^{-2} \Omega$. This technique for fixing boost gauge is used, 
for example, in Ref.~\cite{Korzynski2007}.
In this special family of boost gauges, the extremality is:
\begin{eqnarray}
4 \pi e_0 &=& \oint \left(\epsilon_B{}^C D_C \varpi\right) \left(\epsilon^{BE}D_E \varpi\right)dA\\
&=& - \oint \varpi D^2 \varpi \hspace{1mm} dA\label{e:varpiform}\\
&=& - \oint \varpi \Omega \hspace{1mm} dA\\
&=& - \oint \Omega D^{-2} \Omega \hspace{1mm} dA.
\end{eqnarray}

By fixing to the gauge where $\vec D \cdot \vec \omega = 0$, we have derived 
an expression for extremality entirely computable in terms of boost-invariant 
data on the 2-surface. For this reason, the quantity above can be considered 
a boost-invariant version of the Booth-Fairhurst extremality. It should be 
noted, though, that in general this quantity is {\it not} the same as the 
extremality computed in a boost gauge adapted to a dynamical horizon.

Despite not being adapted to a dynamical horizon, the extremality
defined here has some appealing properties. The one most relevant to
our work here is that it provides a {\it lower bound} on the
extremality, in the space of all choices of boost gauge. In our
preferred family of boost gauges in which Eq.~(\ref{eq:prefGauge}) holds, 
we have:
\begin{equation}
\omega_A = \epsilon_A{}^B D_B \varpi,
\end{equation}
where, again, $\varpi$ is a boost-invariant quantity. In any other boost gauge, 
$\omega_A$ must differ from this by a gradient:
\begin{equation}
\omega_A = \epsilon_A{}^B D_B \varpi + D_A a.
\end{equation}
The extremality computed in this other boost gauge is:
\begin{eqnarray}
4 \pi e &=& \oint \left( \epsilon_A{}^B D_B \varpi + D_A a \right) \left( \epsilon^{BE} D_E \varpi + D^B a \right)dA\\
&=& 4 \pi e_0 + 2 \oint \epsilon_A{}^B D_B \varpi D^A a \hspace{1mm} dA + \oint \lvert\vec D a \rvert^2 \hspace{1mm} dA.
\end{eqnarray}
The second term on the right-hand side vanishes after an integration
by parts, and the final term is manifestly nonnegative. Thus 
\begin{equation}
e \geq e_0,
\end{equation}
with equality only in our preferred family of gauges. Thus, this
measure of extremality can exceed 1 only if the Booth-Fairhurst
extremality measure exceeds 1 in {\it all} boost gauges, including the
one adapted to the dynamical horizon.

As noted above, this measure of extremality is closely related to the 
multipolar structure of the horizon. Closely related formalisms for defining 
such multipolar structure can be found in 
Refs.~\cite{Ashtekar2004,Owen2009,Ashtekar:2013qta}.
In particular, in 
Ref.~\cite{Owen2009}, the eigenfunctions in Eq.~\eqref{e:eigenproblem} are 
taken to define the current multipoles of a horizon 2-surface. Owing to the 
self-adjointness of the operators on both sides of the generalized 
eigenproblem, the eigenfunctions satisfy the condition
\begin{equation}
\oint z_i D^2 z_j \hspace{1mm} dA = 0 \hspace{5mm}\mbox{if $\lambda_i \neq \lambda_j$}
\end{equation}
Hence the eigenfunctions can be normalized in such a way that 
\begin{equation}
\oint z_i D^2 z_j \hspace{1mm} dA = -\delta_{ij}.
\end{equation}
The negative sign is necessary because $D^2$ is a negative-definite operator
in the space of non-constant functions. 
Note also that this is {\it not} the normalization condition used when 
computing spin by approximate Killing vectors. 

Using this orthogonality relation, the potential $\varpi$ can be expanded as 
\begin{equation}
\varpi = \sum_i L_i z_i,
\end{equation}
where the $L_i$ are current multipole moments, defined in 
Ref.~\cite{Owen2009} as 
\begin{equation}
L_i = \oint \varpi D^2 z_i dA = \oint z_i D^2 \varpi \hspace{1mm} dA = \oint z_i \Omega \hspace{1mm}dA.
\end{equation}
Inserting this expansion into Eq.~\eqref{e:varpiform}, we have 
\begin{eqnarray}\label{eq:zeta0}
4 \pi e_0 &=& - \sum_{i, j} L_i L_j \oint z_i D^2 z_j \hspace{1mm} dA\\
&=& \sum_{i, j} L_i L_j \delta_{ij}\\
&=& \sum_i L_i^2.
\end{eqnarray}

\begin{figure}[t]
\includegraphics[width=3.5in]{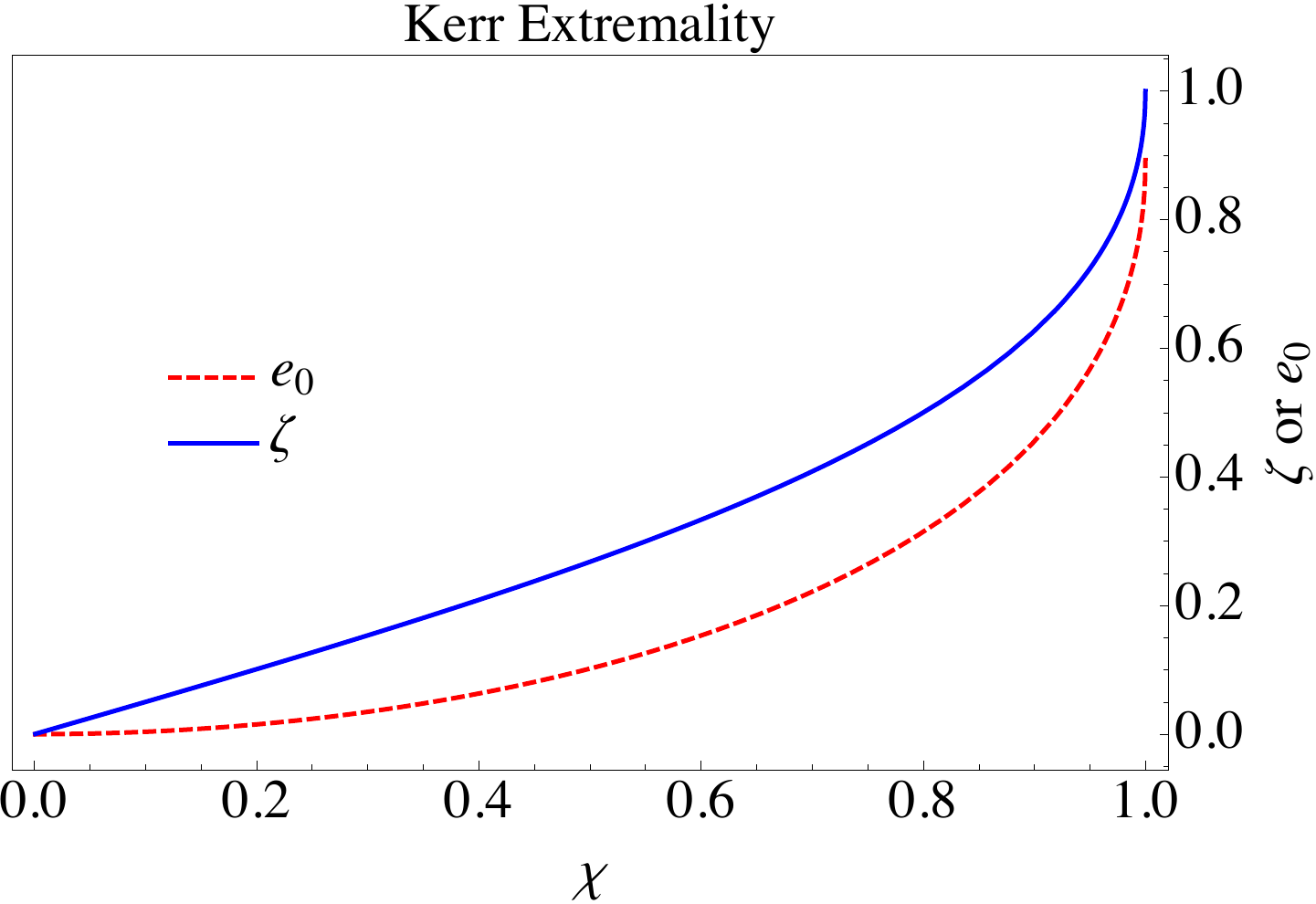}
\caption{The extremality $\zeta$ and extremality lower bound 
$e_0$ for a Kerr black hole of spin $\chi$.
\label{fig:Kerr}}
\end{figure}
Finally, note that the relationship between dimensionless 
spin $\chi$
and our gauge-invariant extremality parameter $e_0$ 
can be calculated 
explicitly on a Kerr horizon. A straightforward calculation gives:
\begin{eqnarray}
e_0^{\rm Kerr}(\zeta) &=& \frac{1}{8 \zeta (1+\zeta^2)}\Big[\Big.
  3\zeta + 8\zeta^3 - 3 \zeta^5   \notag  \\
&& + \left(-3+3\zeta^2+7\zeta^4+\zeta^6\right)\tan^{-1}(\zeta)\Big.\Big]
\end{eqnarray}
where 
\begin{equation}
\zeta \equiv \frac{8 \pi S}{A} = \frac{\chi}{1\pm\sqrt{1-\chi^2}} = \frac{a}{M \pm \sqrt{M^2-a^2}},\label{eq:zetaAndChi}
\end{equation} where the + sign is chosen for subextremal holes.
Figure~\ref{fig:Kerr} compares the extremality measures $e_0$ and $\zeta$ for Kerr horizons.

Note one interesting fact about the lower bound $e_0$: while the
$\chi=0$ case gives $e_0 = 0$, as one would expect, in the extremal
case, $\chi = 1$, $\zeta = 1$, we have $e_0^{\rm Kerr}(\zeta=1) = (4 +
\pi)/8 \approx 0.893$. Thus, in this measure, even extremal Kerr does
not have an extremality of 1.  This could be seen as a weakness of our
choice of boost-gauge, and a reason to prefer the family of boost
gauges favored in Ref.~\cite{BoothFairhurst:2008}, in which extremal
Kerr does indeed have an extremality of 1. On the other hand, in
section~\ref{sec:resultsID}, we will construct BBH initial
  data containing highly distorted, highly dynamical horizons on
which $e_0 > e_0^{\rm Kerr}(\zeta=1)$ while remaining subextremal 
by all of our measures
(cf. Fig.~\ref{fig:OverspunID}). 
Thus, one might intuitively conclude that some amount of
dynamics is necessary for the extremality parameter $e_0$
to approach unity.

\subsection{Simulations}\label{sec:simulationMethods}
We performed the numerical simulations used in this paper with 
the Spectral Einstein Code (SpEC)~\cite{SpECwebsite}. 
We construct~\cite{Pfeiffer2003} 
quasi-equilibrium~\cite{Caudill-etal:2006,Lovelace2008} initial
data  to solve the Einstein constraint 
equations~\cite{York1999} for binaries with low ($\roughly 10^{-4}$)
eccentricity~\cite{Pfeiffer-Brown-etal:2007,Buonanno:2010yk,Mroue:2012kv}.  
In particular, following Ref.~\cite{Lovelace2008}, we base our initial data 
on a weighted superposition of two boosted, spinning Kerr-Schild black 
holes.

We evolve the initial data using a generalized
harmonic 
formulation~\cite{Friedrich1985, Garfinkle2002, Pretorius2005c,Lindblom2006}
of Einstein's equations and damped harmonic 
gauge~\cite{Lindblom2009c,Choptuik:2009ww,Szilagyi:2009qz}.
The adaptively-refined~\cite{Lovelace:2010ne,Szilagyi:2014fna} grid
extends from pure-outflow
excision boundaries conforming to
the shapes of the apparent
horizons~\cite{Scheel2009,Szilagyi:2009qz,Hemberger:2012jz,Ossokine:2013zga}
to an artificial outer boundary, where we enforce constraint-preserving
boundary conditions~\cite{Lindblom2006, Rinne2006, Rinne2007}.
After the holes merge, the grid has only one excision 
boundary~\cite{Scheel2009,Hemberger:2012jz}. 
We use a pseudospectral fast-flow algorithm~\cite{Gundlach1998} 
to find apparent horizons.

Building on the methods of Refs.~\cite{Lovelace:2011nu,Hemberger:2013hsa}, 
we evolve this initial data through inspiral, merger, and ringdown. 
Evolutions with very high black-hole spins are particularly challenging; 
for instance, maintaining pure-outflow 
excision boundaries near the time of merger is especially challenging 
in this case. A companion paper~\cite{Scheel2014}
discusses our methods for 
handling nearly extremal black-hole spins.

\begin{table*}
\begin{tabular}{|lcc|ll|ll|}
\hline
Name & ID & Ref. & $\chi = \chi_{\rm relax}$ & $\zeta_{\rm relax}$ & $\zeta_{\rm final}$ & $\chi_{\rm final}$ \\
\hline
$S^{++}_{0.75}$  & 0175 & - & 0.74994(7) & 0.45136(6) & 0.61982(6) & 0.89558(4) \\
$S^{++}_{0.8}$   & 0155 & \cite{Hemberger:2013hsa} & 0.79987(1) & 0.499868(5) & 0.63910(2) & 0.90753(1)\\
$S^{++}_{0.85}$  & 0153 & \cite{Hemberger:2013hsa} & 0.84983 & 0.55651 & 0.659292 & 0.91909\\
$S^{++}_{0.90}$  & 0160 & \cite{Hemberger:2013hsa} & 0.899737(3) & 0.626370(5) & 0.68047(5) & 0.93021(2) \\
$S^{++}_{0.95}$  & 0157 & \cite{Hemberger:2013hsa} & 0.949586(7) & 0.722940(2) & 0.70275(6) & 0.94085(3)\\
$S^{++}_{0.96}$  & 0176 & - & 0.95956(9) & 0.7488(2) & 0.70733(4) & 0.94291(2)\\
$S^{++}_{0.97}$  & 0158 & \cite{Lovelace:2011nu} & 0.96950 & 0.778672 & 0.712011 & 0.94496 \\
$S^{++}_{0.98}$  & 0172 & \cite{Mroue:2013PRL} & 0.97941(6) & 0.8149(3) & 0.71666(3) & 0.94696(1)\\
$S^{++}_{0.99}$  & 0177 & \cite{Scheel2014} & 0.9893(2) & 0.86306(6) & 0.72135(1) & 0.948930(5)\\
$S^{++}_{0.994}$ & (0178) & \cite{Scheel2014} & 0.9942 & 0.89805 & 0.723761(5) & 0.949924(2)\\
\hline
\end{tabular}
\caption{The numerical binary-black-hole simulations examined in this
  paper. All simulations have equal masses and equal spins aligned with 
the orbital angular momentum. We name 
each simulation $S^{++}_\chi$, where $\chi$ is the spin of the holes 
after the initial relaxation. Each simulation available in the 
SXS catalog~\cite{SXSCatalog} has the label SXS:BBH:{\it ID}. 
(Note that the publicly released simulation
  SXS:BBH:0178 used a newer version of SpEC than the version used for
  $S^{++}_{0.994}$ in this paper. The differences between the two
  versions are on the order of the errors we quote here.)
The ``Ref.'' column lists references describing those simulations that
are presented elsewhere.
The quantity $\zeta_{\rm relax}$ represents $\zeta$ on 
the individual horizons after 
the initial relaxation; $\zeta_{\rm final}$ and $\chi_{\rm final}$ show 
$\zeta$ and $\chi$, respectively, on the 
common horizon at the final 
time. Uncertainties are
estimated by comparing at 
three different numerical resolutions.
There is no uncertainty listed 
for $\chi$ and $\zeta_{\rm relax}$
in case $S^{++}_{0.994}$, because in this case lower resolutions 
only differ from the high resolution after the first 3.5 orbits (see 
Ref.~\cite{Scheel2014} for details).
\label{tab:configurations}}
\end{table*}

We consider a family of BBH simulations with equal masses 
and equal spins aligned with the orbital angular momentum 
(Table~\ref{tab:configurations}). 
Results from most of these have been previously
published~\cite{Lovelace:2011nu,Hemberger:2013hsa,Mroue:2013PRL}.
The simulations $S^{++}_{0.99}$ and $S^{++}_{0.994}$~\cite{Scheel2014}
are new, as are the simulations $S^{++}_{0.75}$ and $S^{++}_{0.96}$. 
We measure the 
approximate-Killing-vector spin $S$ and area $A$ on each apparent horizon. 
The mass $M$ is given by the Christodoulou 
formula\footnote{Note that the dimensionless spin is bounded: 
$\chi = S/M^2 = 2\zeta/(1+\zeta^2)$ satisfies $\chi \leq 1$ 
by construction. In contrast, $\zeta$ is
not trivially bounded in this way.},
\begin{eqnarray}
M^2 \equiv M_{\rm irr}^2 + \frac{S^2}{4 M_{\rm irr}^2},
\end{eqnarray} 
where $M_{\rm irr}\equiv\sqrt{A/16\pi}$ is the irreducible mass.
For most of the simulations, we perform calculations at 
three different numerical resolutions; however, 
for practical reasons\footnote{Simulations $S^{++}_{0.85}$ and 
$S^{++}_{0.97}$ were run with an early version of SpEC,
which made it impractical to calculate $e_0$ and $\zeta$ 
on the lower resolution data.}
we consider only the highest resolution for simulations $S^{++}_{0.85}$ 
and $S^{++}_{0.97}$. This does not significantly 
affect our results or conclusions.

\section{Results}\label{sec:results}
\subsection{Extremality}\label{sec:resultsExtremality}
\begin{figure}[t]
\hspace{-0.1in}\includegraphics[width=3.5in]{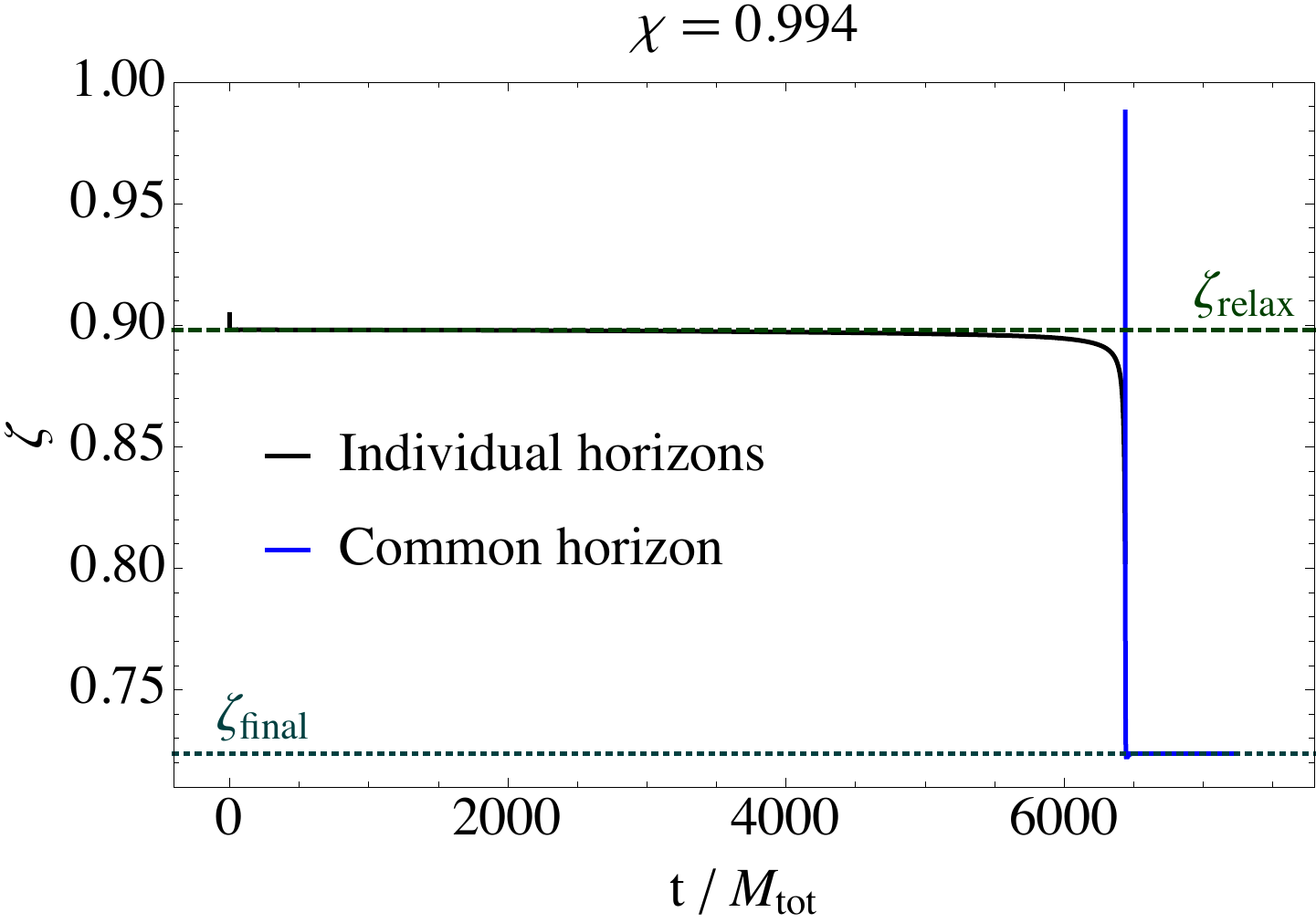}
\caption{
The extremality $\zeta \equiv 8\pi S/A$ as 
a function of time $t/M_{\rm tot}$, where $M_{\rm tot}$ is the 
sum of the holes' initial 
Christodoulou masses, for simulation \run{0.994}
(cf.~Table~\ref{tab:configurations}).
The black and blue curves show the extremality on the 
individual and common apparent horizons, respectively. The 
dashed line is the extremality value after the initial data 
relax to equilibrium, and the dotted line is the final extremality of 
the remnant. \label{fig:AllSpinVsTime}}
\end{figure}

\begin{figure}[t]
\hspace{-0.1in}\includegraphics[width=3.5in]{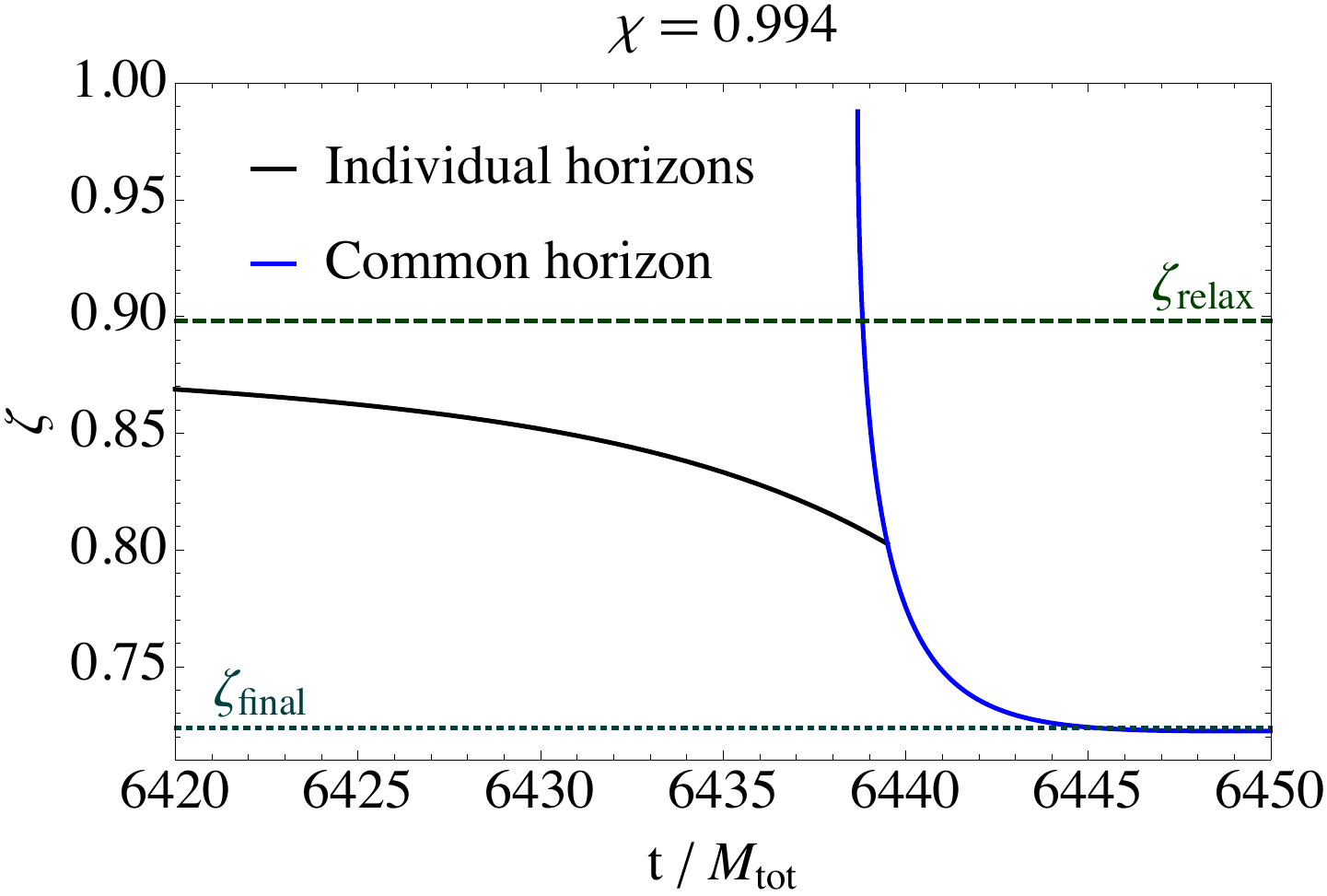}
\caption{
Enlargement of Fig.~\ref{fig:AllSpinVsTime}, zoomed in to show times 
near the formation of the common apparent horizon. Note that the individual 
apparent horizons 
(or more precisely, the individual marginally outer trapped surfaces)
exist even after the common apparent horizon appears, 
until we begin excising the interior of the common apparent horizon; 
by coincidence, at 
that time the common and individual horizons have approximately 
the same extremality 
$\zeta$.
\label{fig:AllSpinVsTimeZoom}}
\end{figure}

In this subsection, we evaluate the extremality measure $\zeta \equiv
8\pi S/A$ for the numerical simulations described in
Sec.~\ref{sec:techniques}. 

In each simulation, after 
the initial data relax and emit spurious gravitational radiation, the
extremality $\zeta$ remains nearly constant until near the time of
merger. Just before merger, $\zeta$ of the individual
horizons decreases as $A$ increases and $S$ decreases; this is caused by 
tidal heating.  The common horizon
initially has a very large $\zeta$ that quickly decreases as the hole
relaxes to the Kerr geometry.  

Figure~\ref{fig:AllSpinVsTime} illustrates this for \run{0.994}, 
and Fig.~\ref{fig:AllSpinVsTimeZoom} zooms 
in on times near the formation of the common apparent horizon.
During the inspiral and late in the ringdown, when the 
holes are nearly Kerr, $\chi$ and $\zeta$ 
are consistent with Eq.~(\ref{eq:zetaAndChi}).
For our higher-spin simulations, including $\run{0.994}$, 
we find that $\zeta$ is closest to unity (but rapidly decreasing) 
on the common horizon just after merger. This motivates us to 
determine with high precision the time when the common
horizon first appears and the value of $\zeta$ at that time.

During a simulation, we typically search for the common horizon
at regular time intervals, starting at
a time when the 
holes' separation becomes 
sufficiently small.  This gives us
a rough measure of when the common horizon forms.  For the simulations
described here, we reran the portion of each simulation near common
horizon formation, searching for a common horizon at each time
step of the simulation.  In each simulation, we find a common horizon
at some earliest time step $t_1$ but not at previous time 
steps.

\begin{figure}[t]
\hspace{-0.1in}\includegraphics[width=3.5in]{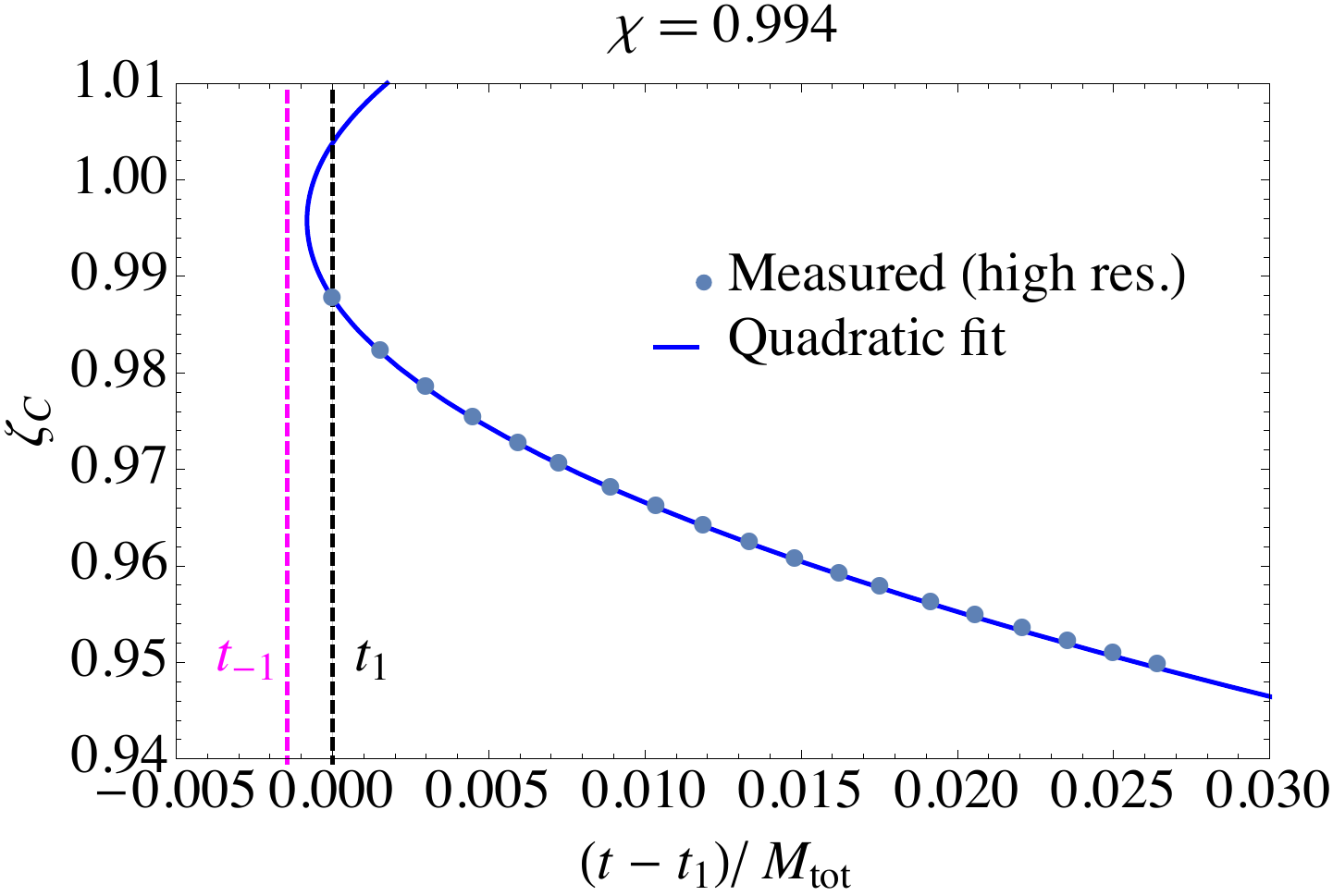}
\caption{
Time $t/M_{\rm tot}$ vs.~the extremality $\zeta_C$ 
of 
the common apparent horizon, for 
the same simulation as in Fig.~\ref{fig:AllSpinVsTime}. For clarity, 
time is shown relative to $t_1/M_{\rm tot}$, the earliest time when we find 
a common apparent horizon (dashed black line). Dots indicate measurements of 
the common apparent horizon at individual simulation time
steps, and the curve indicates a quadratic best fit 
to the 10 earliest of these measurements. The magenta dashed line is the 
latest time step where a common apparent horizon is not found.
\label{fig:LocalMinPlot}}
\end{figure}

We expect, by analogy with Fig.~5 of 
Ref.~\cite{Chu:2010yu}, that the common horizon should first appear as 
a single surface that immediately bifurcates into an inner and 
outer surface.
We find that this expectation holds in the simulations 
we have examined; 
as a concrete example, we plot $\zeta$ 
as a function of time $(t - t_1) / M_{\rm tot}$ 
for simulation $S^{++}_{0.994}$ (Fig.~\ref{fig:LocalMinPlot}), 
where $M_{\rm tot}$ is the sum of the initial Chrsitodoulou masses of the 
individual apparent horizons.
We compute $\zeta$ at the ten earliest time steps where we can find a 
common horizon, and we fit these ten points to a parabola. We
find that the local minimum of this parabola lies at some 
time $t_0$ between 
the earliest time step $t_{1}$ where we can find the horizon 
and the latest step $t_{-1}$ where 
we cannot find it. This gives us confidence that the common horizon 
does appear at the local minimum of this parabola.

\begin{figure}[t]
\hspace{-0.2in}
\includegraphics[width=3.5in]{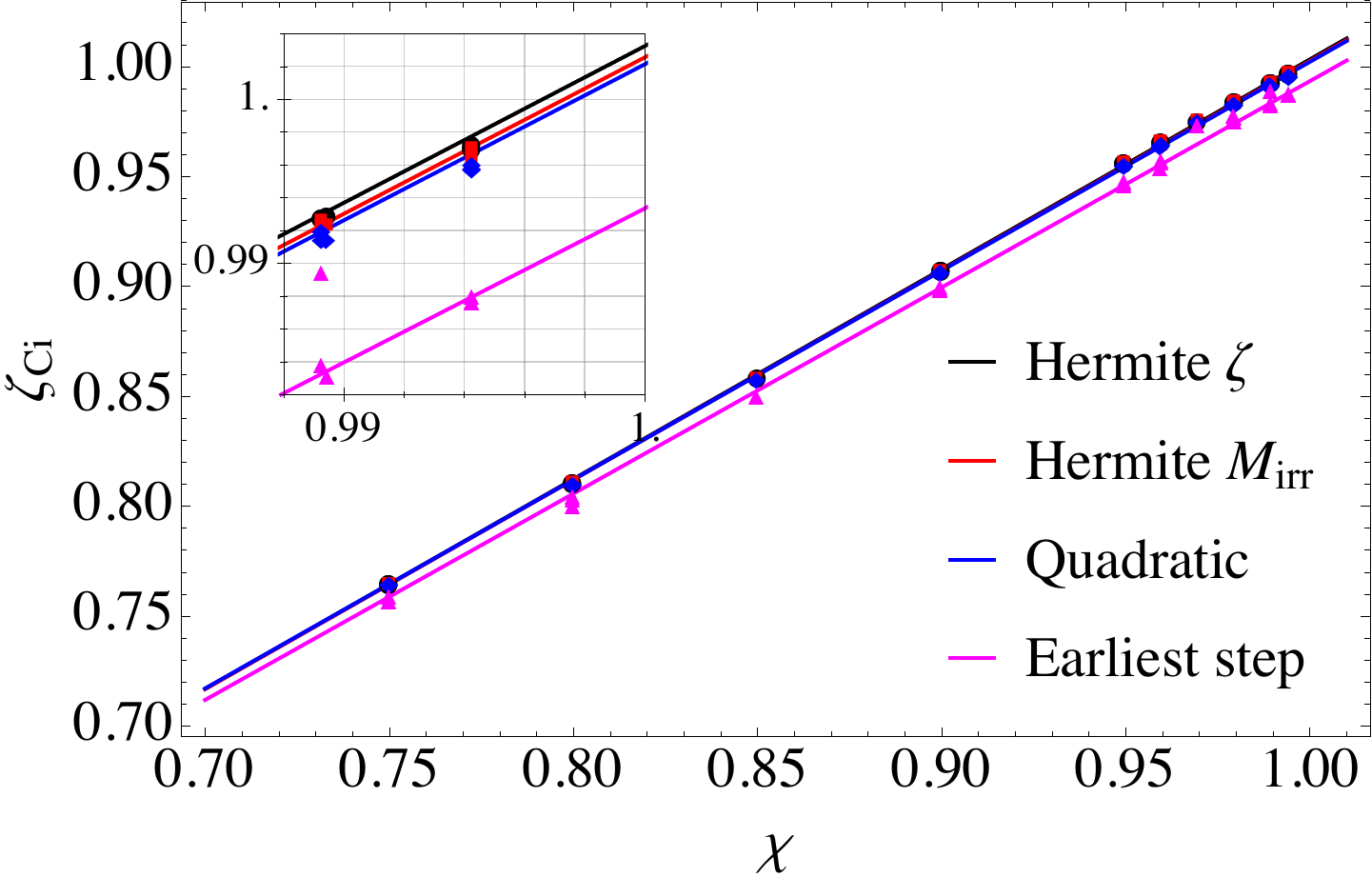}
\caption{
  The extremality $\zeta_{Ci} \equiv 8\pi S/A$ on the common
  apparent horizon, at the moment
  it first appears, for a family of
  merging black holes which initially had  
  equal masses and equal spins of magnitude
  $\chi \equiv S/M^2$ aligned with the orbital angular
  momentum (cf. Table~\ref{tab:configurations}). 
  The spin $\chi$ of the individual
    black holes is measured after the initial data
  have relaxed to equilibrium. Data points (shapes) and linear fits
  to those points (lines) are shown, with different colors indicating
  four different methods of computing $\zeta_{Ci}$. 
  The inset zooms in to
  show the behavior near $\chi=1$. For each of the four different 
extrapolation methods, results are shown for three numerical 
resolutions (except for simulations $S^{++}_{0.85}$ and $S^{++}_{0.97}$); 
differences caused by changing numerical resolution are 
less significant than 
differences caused by different extrapolation methods.
\label{fig:ExtremalityVsChi}}
\end{figure}

Thus we conclude that the common apparent horizon has its
largest value of 
$\zeta$ at time $t_0$, the moment when it first appears. 
For our simulations with nearly extremal spins,
this is the
largest value of 
$\zeta$ on any apparent horizon during the simulation.
In practice, $t_0$ lies between simulation time steps; 
because $\zeta$ is changing so quickly as a function of time, accurate 
estimates of $\zeta$ at $t=t_0$
(which we will refer to as $\zeta_{Ci}$)
require extrapolation. 

We compare four different estimates of $t_0$ and $\zeta_{Ci}$:
\begin{enumerate}
\item Cubic Hermite extrapolation of 
$\zeta(t)$ to find $t_0$ and 
$\zeta_{Ci}$,
\item Cubic Hermite extrapolation of the irreducible mass\footnote{Note that when quantities other than $\zeta$, for instance, the 
irreducible mass $M_{\rm irr}$ of the common horizon, are plotted
versus time, the result is also a parabola as in Fig.~\ref{fig:LocalMinPlot},
so quantities such as $M_{\rm irr}$ can also be used to estimate the
time of common horizon formation $t_0$ as the minimum of this parabola.} 
$M_{\rm irr}(t)$
to find $t_0$, followed by cubic Hermite extrapolation of $\zeta(t)$
to find $\zeta$ at 
$t_0$,
\item Fit a parabola to the 10 earliest measurements of time $t$ as 
a function of the common horizon's extremality $\zeta$, and extrapolate 
using the fitted parabola to find $\zeta_{Ci}$, 
as in Fig.~\ref{fig:LocalMinPlot},
\item Estimate $\zeta_{Ci}$ as the value of $\zeta$ at time $t_1$, the 
earliest simulation time step where we find the common horizon.
\end{enumerate}
In Fig.~\ref{fig:ExtremalityVsChi}, we compare these four estimates of 
$\zeta_{Ci}$ as functions 
of the dimensionless spin $\chi \equiv S/M^2$ 
measured on one of the 
{\it individual}
horizons after the initial relaxation. 
We find an approximately linear relationship. The first three 
estimates, each of which involve extrapolation, give consistent estimates 
of $\zeta_{Ci}$ that are typically
larger than $\zeta(t=t_1)$. 

We conclude from Fig.~\ref{fig:ExtremalityVsChi} 
that the common horizon satisfies the inequality 
$\zeta \equiv 8 \pi S/A \le 1$ 
at all times, even when it first appears. However, for nearly extremal 
initial black-hole spins, the inequality is almost violated at time $t_0$, 
when the common horizon first appears. The approximately linear dependence 
on the initial, relaxed spin $\chi$ suggests that this inequality 
could be slightly violated with even higher initial black-hole spins
than we have simulated.
We speculate that this does not happen in practice, but that 
instead $\zeta_{Ci}$ depends on $\chi$ such that $\zeta_{Ci}$ never 
exceeds unity for any $\chi < 1$. 
Verifying this speculation would require additional simulations beyond SpEC's 
current capabilities.

\subsection{Extremality lower bound}\label{sec:resultsLowerBound}

In the previous subsection, we found that the inequality $\zeta \equiv 8\pi
S/A \le 1$ was satisfied for all of our simulations, but that it was
nearly violated on the common apparent horizons when they first
appeared. This demonstrates that our measure of the horizon spins,
based on approximate Killing vectors, behaves at least somewhat
sensibly, even when the horizons are not very axisymmetric. But does
this necessarily mean that these early common horizons are very close
to extremality? Or, might it be, because the common horizons are
far from axisymmetry, 
that the approximate-Killing-vector spin measure
$S$ does not represent spin in a physically meaningful way?

To address this question, in this subsection we consider an independent 
measure of the extremality of the common horizons, the lower bound 
$e_0$ introduced in Sec.~\ref{sec:extremality}. This
quantity 
is gauge and slicing invariant and does not 
rely on approximate axisymmetry. However, it is only a lower bound, and 
even an extremal Kerr hole has $e_0^{\rm Kerr}(\zeta=1) \approx 0.89$. 

\begin{figure}[t]
\hspace{-0.1in}\includegraphics[width=3.5in]{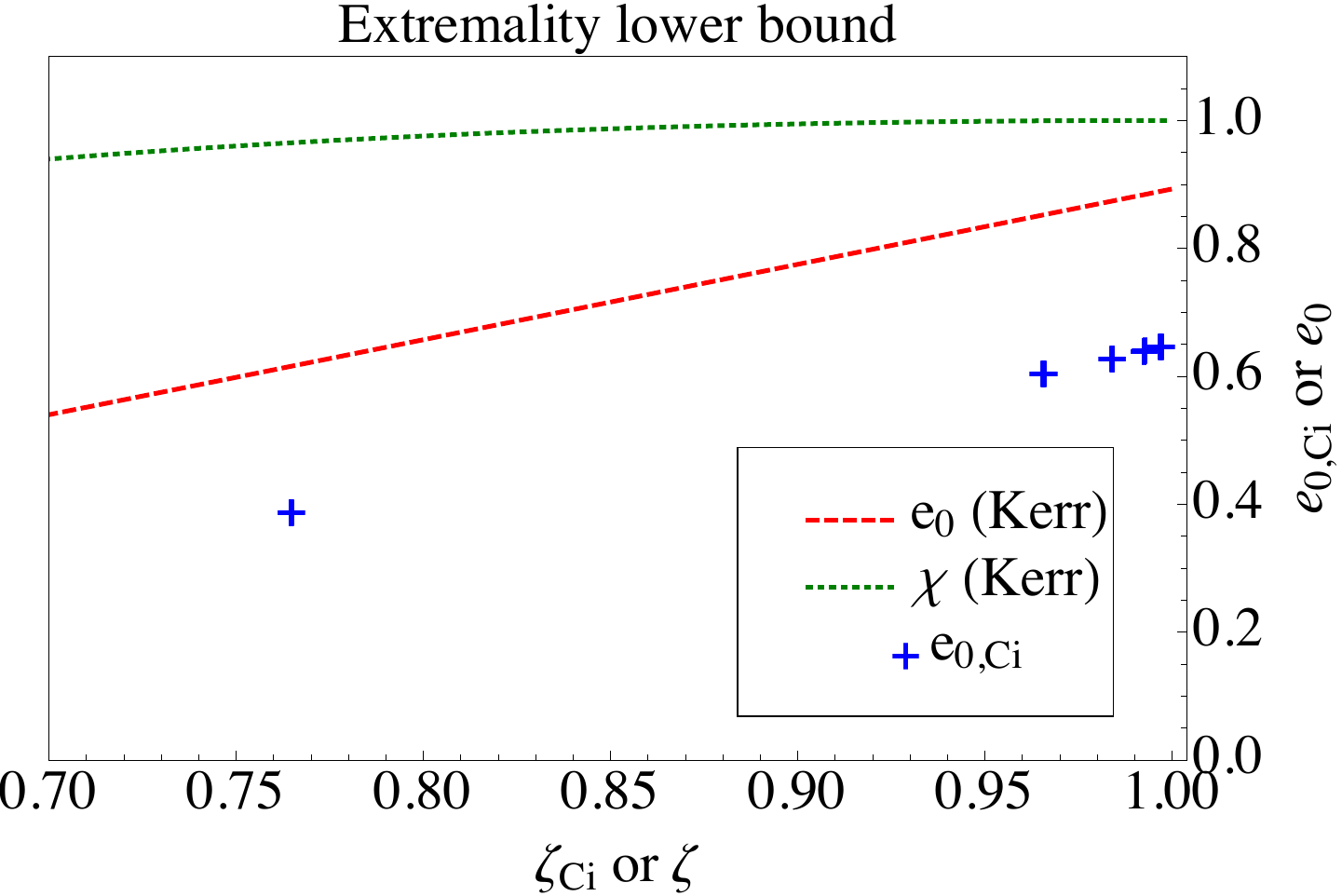}
\caption{The extremality lower bound $e_{0,Ci}$
  [Eq.~(\ref{eq:zeta0})] on the common apparent horizons when they first
  appear in simulations $S^{++}_{0.75}$, $S^{++}_{0.96}$, 
  $S^{++}_{0.98}$, $S^{++}_{0.99}$, and $S^{++}_{0.994}$. 
  The extremality lower bound (dashed)
  $e_0^{\rm Kerr}(\zeta)$ and dimensionless spin
  $\chi\equiv S/M^2=2\zeta/(1+\zeta^2)$ (dotted) for 
  Kerr black holes are also shown.
\label{fig:OwenExtremality}}
\end{figure}

Figure~\ref{fig:OwenExtremality} compares $e_0^{\rm Kerr}(\zeta)$ 
(dashed curve) to $e_0$ measured on the
common apparent horizons when they first appear (data points, shown 
for all resolutions), using cubic Hermite interpolation and extrapolation to 
determine  $e_{0,Ci}$ (method 1 in the previous subsection).
The newborn common apparent horizons have $e_{0,Ci}$  
as large 
as $e_0^{\rm Kerr}(\zeta \approx 0.8)$. While this is
significantly below $e_0^{\rm Kerr}(\zeta = 1)$, it is still 
moderately extremal; note that a Kerr black hole 
with $\zeta \approx 0.8$ has a spin $\chi \approx 0.97$.
This suggests that 
the common horizons are at least moderately close to extremality when 
they first appear, independent of whether the 
approximate-Killing-vector method is a good measure of spin on
apparent horizons that are far from axisymmetric.

\subsection{Extremality lower bound in initial data for merging black holes}
\label{sec:resultsID}
Is it \textit{ever} possible for apparent horizons in numerical
simulations of merging black holes to violate the inequality $e_0
\le 1$?  We observe no such violations in the numerical simulations we
have considered so far. In this subsection, we attempt to
\textit{construct} initial data for merging black holes with apparent
horizons with $e_0 > 1$. 

We follow the methods of Ref.~\cite{Lovelace2008} and the 
references therein to construct
superposed-Kerr-Schild initial data for two equal-mass black holes 
with spins of equal magnitude aligned with the orbital angular momentum. 
In this initial data method, the initial spatial metric $g_{ij}$ is 
proportional to a conformal metric that is a weighted superposition of two 
boosted, spinning Kerr-Schild black holes $\tilde{g}_{ij}^{\rm SKS}$:
\begin{equation}
g_{ij} =  \psi^4 \tilde{g}_{ij}^{\rm SKS}.
\end{equation}
Two regions are excised from the computational 
domain, and boundary conditions are imposed on the excision surfaces. 
One boundary condition enforces that the excision surfaces have 
zero expansion,
{\it i.e.}, that they are marginally outer trapped surfaces. Another 
controls the holes' spins via a parameter $\Omega_r$, which adjusts the 
tangential part of the shift on the boundary. 

\begin{figure}[t]
\hspace{-0.1in}\includegraphics[width=3.5in]{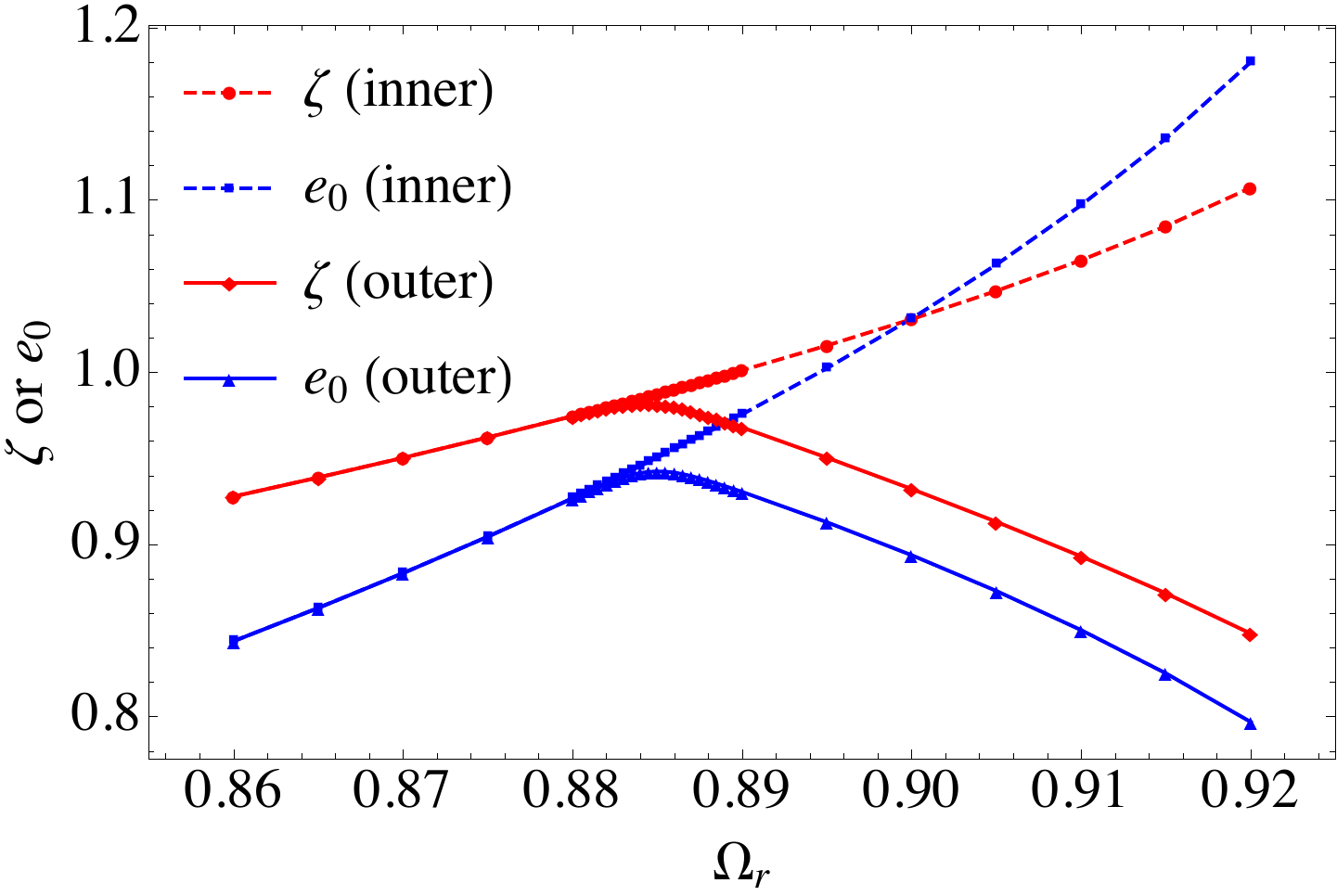}
\caption{The extremality $\zeta$ and extremality lower bound 
$e_0$ for superposed-Kerr-Schild (SKS) initial data for 
binary black holes. Each initial data set yields a binary black hole, 
with equal masses and equal spins aligned with the orbital angular momentum. 
The holes' spins (and thus their extremalities) are controlled by a 
parameter $\Omega_r$ (shown here in arbitrary units).
A boundary condition enforces that the inner, excision surfaces are
marginally outer trapped surfaces; 
at low values of $\Omega_r$, this surface is the 
apparent horizon. For high enough values of $\Omega_r$, the 
excision surface can be superextremal (dashed lines), but in that case, 
the apparent horizon is a larger, subextremal marginally 
outer trapped surface 
that appears (solid lines), 
enclosing the overspun ``inner horizon''. 
\label{fig:OverspunID}}
\end{figure}

Following Ref.~\cite{Lovelace2008}, we construct a family of initial data 
sets with the same conformal metric $\tilde{g}_{ij}^{\rm SKS}$ whose Kerr-Schild
black holes have dimensionless spins
$\tilde{\chi} \equiv \tilde{S}/\tilde{M}^2 = 0.99$.
Each initial data set has a 
different choice of $\Omega_r$. Figure~\ref{fig:OverspunID}
shows $\zeta$ and $e_0$ as a function of $\Omega_r$. 
As in Ref.~\cite{Lovelace2008}, we find that for sufficiently large values 
of $\Omega_r$, we can construct initial data with zero-expansion 
``inner horizons'' that have $\zeta > 1$. These surfaces are 
enclosed by ``outer horizons'' (the apparent horizons) satisfying 
$\zeta < 1$. 

Fig.~\ref{fig:OverspunID}
also shows $e_0$ for these initial data sets. We are able
to construct initial data with inner horizons with extremality lower
bounds $e_0 > 1$. These surfaces are superextremal, but they
are always enclosed by larger apparent horizons that are subextremal
($\zeta < 1$ and $e_0 < 1$). However, note that 
some of the apparent horizons 
{\it do} have $e_0 > e_0^{\rm Kerr}(\zeta=1)$.

\section{Conclusion}\label{sec:conclusions}
We 
have explored the relationship between the area $A$ and 
approximate-Killing-vector spin $S$ for apparent horizons in numerical 
simulations of merging black holes.
In all of the numerical simulations that we have considered
(with initial spins as high as $S/M^2 = 0.994$),
we have observed
no violation of the spin-area inequality $8\pi S / A \leq 1$. This
inequality is nearly violated when the common apparent horizon first appears 
after two holes with nearly extremal spins have merged. 
We cannot rule out small violations of this inequality with even 
larger initial black-hole spins, but we suspect that 
these violations will not occur, even as the initial black-hole 
spins approach unity.

Additionally, 
we have introduced a new, geometric lower bound on the extremality 
of an apparent horizon, $e_0$. This lower bound on the extremality 
is moderately large on the common horizons that come closest to violating 
$8 \pi S/A$, implying that these horizons are at least moderately close 
to extremality. While we have been able to construct initial data with 
marginally outer trapped surfaces 
where $8 \pi S/A > 1$ and $e_0 > 1$, these superextremal 
surfaces 
are always enclosed by subextremal apparent
horizons, with $8\pi S/A < 1$ 
and $e_0 < 1$.

Because we expect that any reasonable definition of spin on
  an apparent horizon should satisfy $8 \pi S / A \leq 1$, our results
  suggest that the approximate-Killing-vector spin might be a 
  reasonable measure on numerical apparent horizons, 
  even when the horizons are far from axisymmetry.
  Future 
  numerical investigations of more generic cases (with unequal masses, 
  unequal spins, and precession) and at even higher black-hole spins 
  (once such simulations are possible) will provide additional tests of
  the inequality $8\pi S/ A \le 1$ using our 
  current method of measuring black-hole spin. 

\begin{acknowledgments}
We are pleased to thank Sergio Dain for the original inspiration for this 
work through his presentation at GR20, Kevin Kuper for assisting in formatting 
some figures, 
and Christian Ott, Saul Teukolsky, and Evan Foley 
for helpful discussions. 

Simulations used in this work
were computed with SpEC~\cite{SpECwebsite}. Figures were prepared 
and some calculations 
were performed using Mathematica.
This work was supported in part by 
the Sherman Fairchild Foundation; NSF
grants PHY-1306125 and AST-1333129 at Cornell, NSF grants
PHY-1440083 and AST-1333520 at Caltech, and 
NSF grant PHY-1307489 at California State University Fullerton; 
a 2013--2014 California State University Fullerton 
Junior Faculty Research Grant;  
and 
NSERC of Canada, the Canada Chairs Program, 
and the Canadian Institute for Advanced Research.
Computations
were performed on the Zwicky cluster at Caltech, which is supported by
the Sherman Fairchild Foundation and by NSF award PHY-0960291; on the
NSF XSEDE network under grant TG-PHY990007N; 
on the Orca cluster supported by  
NSF award NSF-1429873, by the Research Corporation for Science Advancement, 
and by California State University Fullerton; 
and on the GPC
supercomputer at the SciNet HPC Consortium~\cite{scinet}. SciNet is
funded by: the Canada Foundation for Innovation under the auspices of
Compute Canada; the Government of Ontario; Ontario Research
Fund--Research Excellence; and the University of Toronto. 
\end{acknowledgments}

\bibliography{References/References}

\end{document}